\documentclass[
secnumarabic, amssymb, amsmath, nofootinbib,tightenlines,showpacs,nobibnotes, aps, prl]{revtex4}
\usepackage{amsfonts}
\usepackage{docs}%
\usepackage{bm}%
\usepackage[colorlinks=true,linkcolor=blue]{hyperref}%
\usepackage{epsfig}
\usepackage{graphicx}
\usepackage{subfigure}
\usepackage{dcolumn}
\usepackage{bm}

\begin{document}

\title{Generalized quantization condition in topological insulator}

\author{Yuanpei Lan$^1$}
\author{Shaolong Wan$^1$}
\altaffiliation{Corresponding author} \email{slwan@ustc.edu.cn}
\author{Shou-Cheng Zhang$^2$}
\affiliation{$^1$Department of Modern Physics, University of
Science and Technology of China, Hefei, 230026, {\bf P. R. China}
\\
$^2$Department of Physics, Stanford University, Stanford,
CA94305-4045, {\bf USA}}

\date{\today}

\begin{abstract}
The topological magnetoelectric effect (TME) is the fundamental
quantization effect for topological insulators
in units of the fine structure constant
$\alpha$. In [Phys. Rev. Lett. 105, 166803(2010)],
a topological quantization condition of the TME
is given under orthogonal incidence of the optical beam, in which the wave length of the
light or the thickness of the TI film must be tuned to some commensurate values. This
fine tuning is difficult to realize experimentally. In this article, we give manifestly $SL(2,\mathbb{Z})$
covariant expressions for Kerr and Faraday angles at oblique
incidence at a topological insulator thick film. We obtain a
generalized quantization condition independent of
material details, and propose a more easily realizable optical
experiment, in which only the incidence angle is tuned, to directly measure the topological quantization
associated with the TME.
\end{abstract}


\maketitle


\section{Introduction}
Recently, the time-reversal invariant topological insulator (TI)
has been investigated extensively \cite{qi1, hasan, moore1, qi2}.
The concept of the topological insulator can be defined both
within the topological field theory(TFT)\cite{qi3} and topological
band theory(TBT)\cite{kane1, fu1, moore2}. The TFT is generally
valid for interacting systems and describing a quantized
magnetoelectric response, i.e. topological magnetoelectric
effect(TME) \cite{qi3}. Moveover, the TFT reduces exactly to the
TBT in the non-interacting limit \cite{wang1}. TME in topological
insulator is defined as a magnetization induced by an electric
field or a charge polarization induced by a magnetic field, this
effect is essentially a surface effect although it looks like a
bulk response. The most important feature of TME is the
quantization of the magnetoelectric coefficient
$\partial{M}/\partial{E}$ or $\partial{P}/\partial{B}$ in odd
units of the fine structure constant \cite{qi3, Essin1} However,
since the standard Maxwell term has the same scaling dimension as
the topological term, non-universal materials constants such as
$\varepsilon$ and $\mu$ can mask the exact quantization of the
TME\cite{qi3,joseph}. Hence, in order to observe the topological
quantization of the TME, one should design an experiment to remove
the dependence of the non-topological material properties of the
TI such as $\varepsilon$ and $\mu$. Recently new proposals have
been made to remove the dependence on those non-topological
material constants \cite{joseph, tse1}. In Ref.\cite{joseph}, one
considers a thick film of topological insulator with two surfaces,
with vacuum on one side and a substrate on the other. With light
rays normally incident at the slab, the combination of Kerr and
Faraday angles measured at reflectivity minima or maxima can
directly give topological quantization of the TME. In a more
special case, Ref.\cite{tse1} considers a topological insulator
thin film weakly exchange coupled to ferromagnet and finds that in
the low frequency limit, both the Faraday rotation and the Kerr
rotation are universal, dependent only on the vacuum fine
structure constant. But these quantization condition apply only to
orthogonal incidence, and the thickness of TI film or the
frequency of the incident light must be tuned to specific values
which may be difficult in practical experiments.

In this paper, we consider the general case of oblique incidence, and
obtain a generalized topological quantization condition, in which
the reflectivity minima can be easily realized by changing
incidence angle rather than by tuning the film thickness.
This article is organized as follows. In Sec. II, we present Kerr
and Faraday rotation at a unique interface with oblique incidence.
In Sec. III, Kerr and Faraday rotation on a topological insulator
thick film at oblique incidence are given. In Sec. IV, Generalized
quantization condition in TME is obtained, which is easily realizable
optical experiments. The conclusion is given
in Sec. V.

\section{Kerr and Faraday rotation at a unique interface with oblique incidence}

According to TFT, the effective Lagrangian of a  topological
insulator is given by \cite{qi1, qi3}
\begin{eqnarray}
\mathcal{L} = \mathcal{L}_{0} + \mathcal{L}_{\theta} =
\frac{1}{8\pi} (\varepsilon \vec{E}^{2} - \frac{1}{\mu}
\vec{B}^{2}) + \frac{\theta}{2\pi} \frac{\alpha}{2\pi} \vec{E}
\cdot \vec{B} \label{2.1}
\end{eqnarray}
where $\vec{E}$ and $\vec{B}$ are the electric and magnetic
fields, $\varepsilon$ and $\mu$ are the dielectric constant and
magnetic permeability, respectively, $\theta$ is the axion angle
\cite{wilczek}, and $\alpha$ is the fine structure constant. Under
the time-reversal transformation, $\vec{E}\longrightarrow
\vec{E}$, $\vec{B}\longrightarrow -\vec{B}$, so  for a periodic
system, there are only two values of $\theta$, namely $\theta = 0$
and $\theta = \pi$ (modulo $2\pi$), that give a time-reversal
symmetric theory.

In the framework of $SL(2,\mathbb{Z})$ electric-magnetic duality
symmetry, the constitutive relation can be written in the compact
form \cite{karch1}
\begin{eqnarray}
\left(\begin{array}{c}
\vec{D}\\
2\alpha\vec{B}
\end{array}\right) = \mathcal{M}\left(\begin{array}{c}
2 \alpha \vec{E}\\
\vec{H}
\end{array}\right) \label{2.2}
\end{eqnarray}
with
\begin{eqnarray}
\mathcal{M} = \frac{1}{c}\frac{2\alpha}{c\varepsilon}
\left(\begin{array}{cc}
\frac{\theta^{2}}{4\pi^2} +(\frac{c \varepsilon}{2 \alpha})^2 &  \frac{\theta}{2 \pi}\\
\\
\frac{\theta}{2\pi} & 1
\end{array}
\right) \label{2.3}
\end{eqnarray}
where $\mathrm{det}(\mathcal{M}) = \frac{1}{c^2}$ is duality
invariant under a general $SL(2,\mathbb{Z})$ transformation. On
the other hand, in the presence of a non-trivial $\theta$ angle
Maxwell's equations still allow propagating wave solutions with
$\omega = ck$, all fields orthogonal to the direction of
propagation, and
\begin{eqnarray}
\left(\begin{array}{c}
2 \alpha \vec{E}\\
\vec{H }
\end{array}
\right) = c \hat{k} \times \left(\begin{array}{cc}
0 & -1 \\
1 & 0
\end{array} \right) \left(\begin{array}{c}
\vec{D} \\
2\alpha\vec{B}
\end{array} \right) \label{2.4}
\end{eqnarray}
where $\hat{k} = \frac{\vec{k}}{\mid \vec{k} \mid}$ is the unit
wave vector. That is, $\vec{E} \perp \vec{B}$ and $\vec{D} \perp
\vec{H}$. However the 2-bein defined by  $\hat{D}$, $\hat{H}$ is
no longer aligned with the $\hat{E}$, $\hat{B}$ 2-bein due to the
topological $\theta$ term, which is distinguished from the classic
electrodynamics \cite{jackson}.

For simplicity, we denote
\begin{eqnarray} \vec{\mathcal{E}} =
\left(\begin{array}{c}
2 \alpha \vec{E} \\
\vec{H}
\end{array} \right),~~~~~
\vec{\mathcal{D}} = \left(\begin{array}{c}
\vec{D} \\
2 \alpha \vec{B}
\end{array} \right) \label{2.5}
\end{eqnarray}

From equation(\ref{2.2})-(\ref{2.4}), it can obtained the
following relationship
\begin{eqnarray}
\vec{\mathcal{E}} = c \hat{k} \times (-i\sigma_2) \mathcal{M}
\vec{\mathcal{E}} \label{2.6}
\end{eqnarray}

Since $\vec{E}$ and $\vec{H}$ are not independent for a wave
solution, we can express $\vec{\mathcal{E}}$ as
\begin{eqnarray}
\vec{\mathcal{E}} = \left( \begin{array}{c}
2 \alpha \vec{E} \\
c \varepsilon \hat{k} \times \vec{E} - \frac{2 \alpha \theta}{2
\pi} \vec{E}
\end{array} \right) \label{2.7}
\end{eqnarray}
These results contain all the electric-magnetic
relations of the TI in compact forms.

The Kerr and Faraday rotations at orthogonal incidence at the
interface between two materials with different $\varepsilon$,
$\mu$ and $\theta$ have been calculated in Ref.\cite{qi3}. As a
generalization, we  now consider the corresponding problem in the
case of oblique incidence, i.e. the angle of incidence is not
restrict to zero.

\begin{figure}
\includegraphics[scale=0.6]{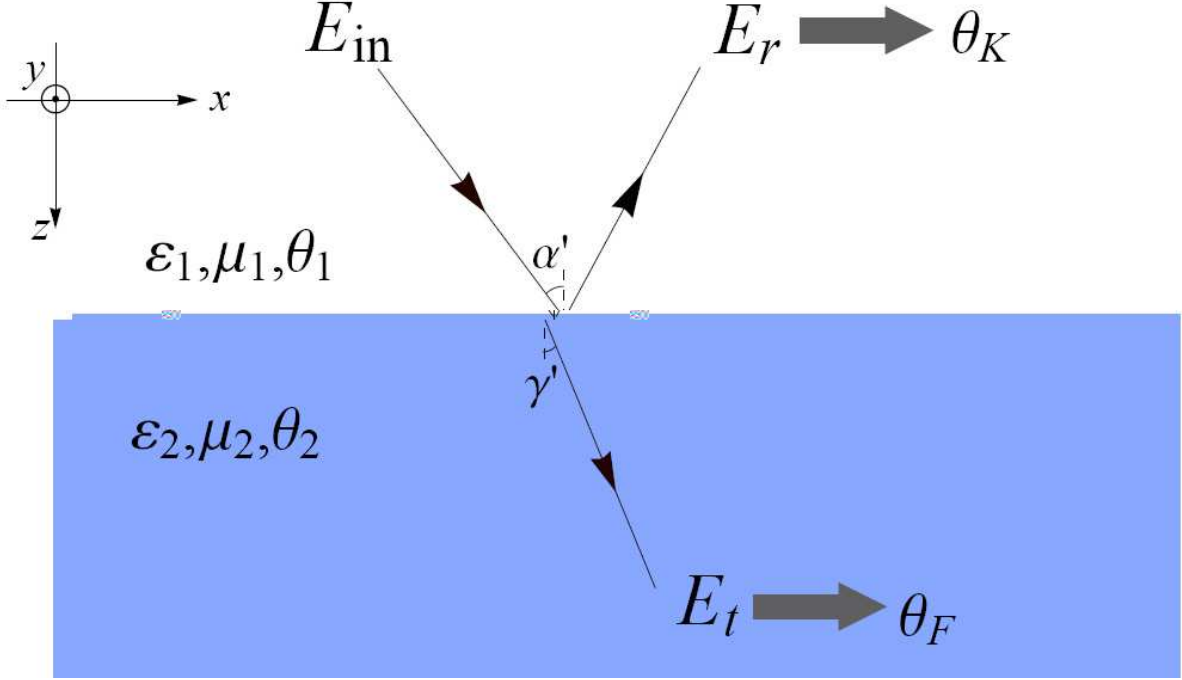}
\caption{Kerr and Faraday rotation at a single surface with
oblique incidence} \label{fig.1}
\end{figure}

Since Snell's law is unmodified even in the presence of a jump in
the axion angle $\theta$, the main task for us is to calculate the
components of the reflected and transmitted electric fields.
Consider light ray shown in Fig.[\ref{fig.1}], incident at an
certain angle $\alpha'$ at a plane interface separating two such
materials mentioned before, and linearly polarized in the y
direction $\vec{E}_{in} = E_{in} \hat{y}$. The wave vector of
incidence, reflection and refraction are expressed as
\begin{eqnarray}
\vec{k}_{in} = k_{in}(\sin\alpha', 0, \cos\alpha'), ~~\vec{k}_r =
k_r(\sin\alpha', 0, -\cos\alpha'),~~ \vec{k}_t = k_t(\sin\gamma',
0, \cos\gamma') \label{2.8}
\end{eqnarray}
Because $\hat{k_i} \cdot{\vec{\mathcal{E}_i}} = 0$, the components
of $\vec{\mathcal{E}}$ are not independent: $\mathcal{E}_r^x =
\cot\alpha' \mathcal{E}_r^z ,  ~~ \mathcal{E}_t^x = - \cot\gamma'
\mathcal{E}_t^z$. Then we can obtain
\begin{eqnarray}
\tan\theta_K = \frac{\vec{E_r} \cdot (\hat{k_r} \times
\hat{y})}{\vec{E_r} \cdot \hat{y}} = \frac{\cos\alpha' E_r^x +
\sin\alpha' E_r^z}{E_r^y} = \frac{1}{\sin\alpha'}
\frac{E_r^z}{E_r^y} \label{2.9}
\end{eqnarray}
\begin{eqnarray}
\tan\theta_F = \frac{\vec{E_t} \cdot (\hat{y} \times
\hat{k_t})}{\vec{E_t} \cdot \hat{y}} = \frac{\cos\gamma' E_t^x -
\sin\gamma' E_t^z}{E_t^y} = - \frac{1}{\sin\gamma'}
\frac{E_t^z}{E_t^y} \label{2.10}
\end{eqnarray}

In this case, all the fields no longer parallel to the interface,
so the boundary conditions are continuity of the
parallel component of $\mathcal{\vec{E}} = \left( \begin{array}{c} 2 \alpha \vec{E} \\
\vec{H} \end{array} \right)$ as well as the vertical component of
$\mathcal{\vec{D}} = \left( \begin{array}{c} \vec{D} \\ 2 \alpha
\vec{B} \end{array} \right)$, which is different from and more
complicated than the case of normal incidence. Two relations can
be obtained
\begin{eqnarray}
\hat{z} \times (\vec{\mathcal{E}}_{in} + \vec{\mathcal{E}}_{r}) =
\hat{z}\times\vec{\mathcal{E}}_{t}  \label{2.11} \\
\hat{z} \cdot (\vec{\mathcal{D}}_{in} + \vec{\mathcal{D}}_{r}) =
\hat{z}\cdot\vec{\mathcal{D}}_{t} \label{2.12}
\end{eqnarray}

Using Eq.(\ref{2.6}), we get the following equation,
\begin{eqnarray}
\mathcal{T}_{12}[\hat{z} \times (\hat{k}_{in} \times
\vec{\mathcal{E}}_{in} + \hat{k}_{r} \times
\vec{\mathcal{E}}_{r})] = \hat{z} \times (\hat{k}_t \times
\vec{\mathcal{E}}_t) \label{2.13}
\end{eqnarray}
where we define a transfer matrix as $\mathcal{T}_{ij} =
\frac{c_i}{c_j} \mathcal{M}_j^{-1} \mathcal{M}_i$ = $\frac{4
\alpha^2}{Y_1 Y_2} \left( \begin{array}{cc}
\frac{Y_i^2}{4 \alpha^2} - \frac{\theta_i (\theta_j -\theta_i)}{4 \pi^2} & - \frac{\theta_j - \theta_i}{2\pi} \\
\\
- \frac{\theta_j}{2 \pi} \frac{Y_i^2}{4 \alpha^2} +
\frac{\theta_i}{2 \pi}\frac{Y_j^2}{4 \alpha^2} + \frac{\theta_i
\theta_j(\theta_j - \theta_i)}{8 \pi^3} & \frac{Y_j^2}{4 \alpha^2}
+ \frac{\theta_j (\theta_j - \theta_i)}{4 \pi^2}
\end{array} \right)$ and $Y_i = \sqrt{\frac{\varepsilon_i}{\mu_i}}$ denotes the admittance of material $i$.

From the constitutive relation, Eq.(\ref{2.12}) is equivalent to
\begin{eqnarray}
\frac{c_2}{c_1} \mathcal{T}_{12}\hat{z} \cdot
(\vec{\mathcal{E}}_{in} + \vec{\mathcal{E}}_{r}^z) = \hat{z} \cdot
\vec{\mathcal{E}}_t \label{2.14}
\end{eqnarray}

We choose the independent equations of y- and z-components from
Eqs.(\ref{2.11})-(\ref{2.14}),
\begin{eqnarray}
&&\frac{\cot\alpha'}{\cot\gamma'}(\mathcal{E}_{in}^z -
\mathcal{E}_r^z) = \mathcal{E}_t^z \label{2.15a} \\
&&\mathcal{E}_{in}^y + \mathcal{E}_r^y = \mathcal{E}_t^y \label{2.15b} \\
&&\frac{c_2}{c_1} \mathcal{T}_{12} (\mathcal{E}_{in}^z +
\mathcal{E}_r^z) = \mathcal{E}_t^z \label{2.15c} \\
&&\frac{\cos\alpha'}{\cos\gamma'} \mathcal{T}_{12}
(\mathcal{E}_{in}^y - \mathcal{E}_r^y) = \mathcal{E}_t^y
\label{2.15d}
\end{eqnarray}
where $\mathcal{E}_{in}^z = E\left( \begin{array}{c}
0 \\
Y_1{\sin\alpha'}
\end{array} \right)$ and  $\mathcal{E}_{in}^y = 2 \alpha{E} \left( \begin{array}{c}
1 \\
-\frac{\theta_1}{2\pi} \end{array} \right)$.

From above, we obtain the $\vec{\mathcal{E}}_r$ and
$\vec{\mathcal{E}}_t$ simultaneously, here we give
$\vec{\mathcal{E}}_r$ only,
\begin{eqnarray}
\mathcal{E}_r^z = (1 + \frac{\cos\gamma'}{\cos\alpha'}
\mathcal{T}_{12})^{-1} (1 - \frac{\cos\gamma'}{\cos\alpha'}
\mathcal{T}_{12}) \mathcal{E}_{in}^z  \label{2.16} \\
\mathcal{E}_r^y = - (1 + \frac{\cos\alpha'}{\cos\gamma'}
\mathcal{T}_{12})^{-1} (1 - \frac{\cos\alpha'}{\cos\gamma'}
\mathcal{T}_{12}) \mathcal{E}_{in}^y \label{2.17}
\end{eqnarray}

At last we get the polarization plane of reflected electric field  rotated by an angle $\theta_K$ with
\begin{eqnarray}
\tan\theta_K = \frac{4 \alpha}{Y_2} \frac{\theta_2 - \theta_1}{2
\pi} \frac{\frac{\cos\alpha'}{\cos\gamma'}}{1 +
\frac{\cos\alpha'}{\cos\gamma'} (\frac{Y_2}{Y_1} -
\frac{Y_1}{Y_2}) - \frac{\cos^2\alpha'}{\cos^2\gamma'}} +
o(\alpha^2) \label{2.18}
\end{eqnarray}
In the limit of orthogonal incidence, i.e. $\alpha' \rightarrow
0$, we obtain $\tan\theta_K = 4 \alpha \frac{\theta_2 -
\theta_1}{2 \pi} \frac{Y_1}{Y_2^2 - Y_1^2} + o(\alpha^2)$, which
is a direct consequence of a jump in $\theta$, and is perfect
agreement with the result in Ref. \cite{qi3, karch1}.

\section{Kerr and Faraday rotation on a topological insulator thick film at oblique incidence }
\begin{figure}
\includegraphics[scale=0.6]{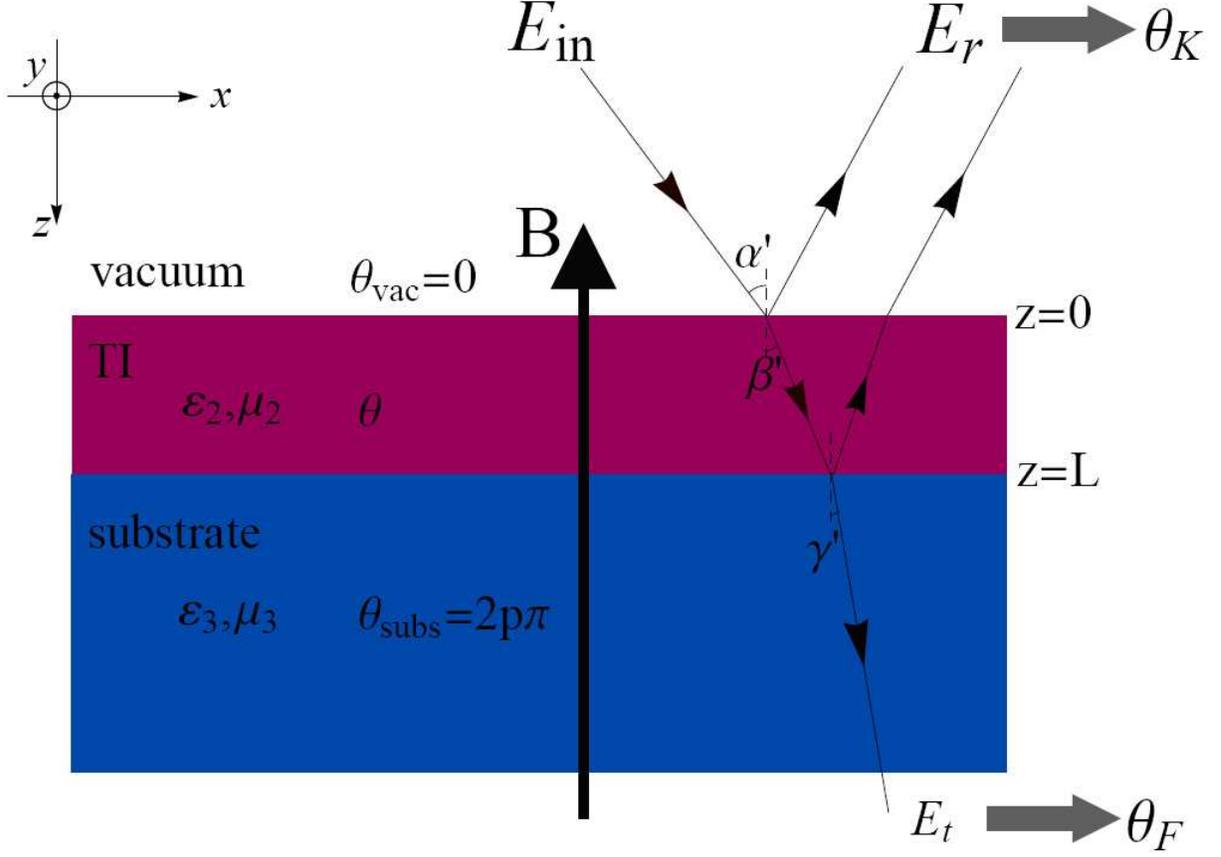}
\caption{Kerr and Karaday rotations at oblique incidence at a TI
thick film} \label{fig.2}
\end{figure}

Now, as shown in Fig.[\ref{fig.2}], we consider a TI thick film of
thickness L with optical constants $\varepsilon_2, ~\mu_2$ and
axion angle $\theta$ deposited on a topological trivial insulating
substrate with optical constant $\varepsilon_3, ~\mu_3$. The substrate
is characterized by an axion angle $\theta_{sub}=2p\pi$, where $p$ is
an integer. Generally, only the change of the $\theta$ angle across
an interface is physically important, which defines the Hall conductance
of the interface\cite{qi3}. In our setup, the two interfaces are defined by the
difference of $\theta$ at the upper interface
and $\theta_{sub}-\theta$ at the low interface. Therefore, $p$ is a measure
of the total Hall conductances of both surfaces. The
vacuum outside the TI has $\varepsilon = \mu = 1$ and trivial
axion angle $\theta_{vac}=0$. Because the reflectivity at
TI/vacuum interface is quite large, the effect of multiple
reflections should be considered. For the oblique incidence, shown
in Fig.[\ref{fig.2}], with the angle of incidence is $\alpha'$, we
have
\begin{equation}
\vec{\mathcal{E}}(\vec{r},t) = \left\{\begin{array}{cc}
\vec{\mathcal{E}}_1^+ e^{ik_{1} (x \sin\alpha' + z \cos\alpha') -
i \omega{t}} + \vec{\mathcal{E}}_1^- e^{i k_1 (x \sin\alpha' - z
\cos\alpha') - i \omega{t}}, & z > 0 \\
\\
\vec{\mathcal{E}}_2^+ e^{ik_{2} (x \sin\beta' + z \cos\beta') - i
\omega{t}} + \vec{\mathcal{E}}_2^- e^{i k_2 (x \sin\beta' -
z\cos\beta') -i \omega{t}}, & 0 < z < L \\
\\
\vec{\mathcal{E}}_3^+ e^{i k_{3} (x \sin\gamma' + z \cos\gamma') -
i \omega{t}}, & z > L
\end{array} \right. \label{3.19}
\end{equation}

For the oblique incidence case, the boundary conditions boundary
condition at $z = 0 $ and $ z = L $ are the continuity of the
parallel component of  $\mathcal{\vec{E}}$  as well as the
vertical component of $\mathcal{\vec{D}}$, using the fact that
$k_i \sin\phi_i = \omega \frac{\sin\phi_i}{c_i} = const$, we get
\begin{eqnarray}
&&\hat{z} \times (\vec{\mathcal{E}}_1^{+} + \vec{E}_{1}^{-}) =
\hat{z} \times (\vec{\mathcal{E}}_2^{+} + \vec{\mathcal{E}}_2^{-})
\label{3.20} \\
&&\hat{z} \times (\vec{\mathcal{E}}_2^+ e^{i k_2 L \cos\beta'} +
\vec{\mathcal{E}}_2^- e^{- i k_2 L \cos\beta '}) = \hat{z} \times
\vec{\mathcal{E}}_3^+ e^{i k_3 L \cos\gamma'} \label{3.21} \\
&&\hat{z} \cdot (\vec{\mathcal{D}}_1^+ + \vec{\mathcal{D}}_1^-) =
\hat{z} \cdot (\vec{\mathcal{D}}_2^+ + \vec{\mathcal{D}}_2^-)
\label{3.22} \\
&&\hat{z} \cdot(\vec{\mathcal{D}}_2^+ e^{i k_2 L \cos\beta'} +
\vec{\mathcal{D}}_2^- e^{- i k_2 L \cos\beta'}) = \hat{z} \cdot
\vec{\mathcal{D}}_3^+ e^{i k_3 L \cos\gamma'} \label{3.23}
\end{eqnarray}

Similar to the method in section II, using Eq.(\ref{2.2}) and
(\ref{2.6}), and choosing the independent z and y components of
$\mathcal{E}$, we acquire two sets of equations.

For z component, we get
\begin{eqnarray}
&&\frac{\cot\alpha'}{\cot\beta'}(\mathcal{E}_{1z}^{+} -
\mathcal{E}_{1z}^{-}) = \mathcal{E}_{2z}^{+} -
\mathcal{E}_{2z}^{-} \label{3.24} \\
&&\frac{c_2}{c_1} \mathcal{T}_{12} (\mathcal{E}_{1z}^{+} +
\mathcal{E}_{1z}^{-}) = \mathcal{E}_{2z}^{+} \mathcal{E}_{2z}^{-}
\label{3.25} \\
&&\frac{\cot\beta'}{\cot\gamma'} (\mathcal{E}_{2z}^{+} e^{i k_2 L
\cos\beta'} - \mathcal{E}_{2z}^- e^{- i k_2 L \cos\beta'}) =
\mathcal{E}_{3z}^+ e^{i k_3 L \cos\gamma'} \label{3.26} \\
&&\frac{c_3}{c_2} \mathcal{T}_{23} (\mathcal{E}_{2z}^{+} e^{ i k_2
L \cos\beta'} + \mathcal{E}_{2z}^- e^{- i k_2 L \cos\beta'}) =
\mathcal{E}_{3z}^+ e^{i k_3 L \cos\gamma'} \label{3.27}
\end{eqnarray}

For y component, we get
\begin{eqnarray}
&&\mathcal{E}_{1y}^+ + \mathcal{E}_{1y}^- = \mathcal{E}_{2y}^+ + \mathcal{E}_{2y}^- \label{3.28} \\
&&\frac{\cos\alpha'}{\cos\beta'} \mathcal{T}_{12} (\mathcal{E}_{1y}^+ - \mathcal{E}_{1y}^-) = \mathcal{E}_{2y}^+ - \mathcal{E}_{2y}^- \label{3.29} \\
&&\mathcal{E}_{2y}^{+} e^{i k_2 L \cos\beta'} + \mathcal{E}_{2y}^- e^{- i k_2 L \cos\beta'} = \mathcal{E}_{3y}^+ e^{i k_3 L \cos\gamma'} \label{3.30} \\
&&\frac{\cos\beta'}{\cos\gamma'} \mathcal{E}_{2y}^{+} e^{i k_2 L
\cos\beta'} - \mathcal{E}_{2y}^- e^{- i k_2 L \cos\beta'} =
\mathcal{E}_{3y}^+ e^{i k_3 L \cos\gamma'} \label{3.31}
\end{eqnarray}

After some algebra, we obtain
\begin{eqnarray}
\mathcal{E}_{1z}^- = V_K
\mathcal{E}_{1z}^+,~~~~~~e^{ik_3L\cos\gamma'}\mathcal{E}_{3z}^-=U_F\mathcal{E}_{1z}^+
\label{3.32} \\
\mathcal{E}_{1y}^- = V_K' \mathcal{E}_{1y}^+, ~~~~~~e^{i k_3 L
\cos\gamma'} \mathcal{E}_{3y}^+ = U_F' \mathcal{E}_{1z}^+
\label{3.33}
\end{eqnarray}
where
\begin{eqnarray}
&&U_F = \frac{c_3}{c_2}[\mathbb{I} +
\frac{\cos\gamma'}{\cos\beta'} \mathcal{T}_{23} \mathcal{Q}_{12}^*
(\mathcal{P}_{12}^{*})^{-1}]^{-1} \mathcal{T}_{23}
[\mathcal{Q}_{12} + \mathcal{Q}_{12}^{*} (\mathcal{P}_{12}^*)^{-1}
\mathcal{P}_{12}] \label{3.34} \\
&&U_F' = \frac{\cos\beta'}{\cos\gamma'} [\mathbb{I} +
\frac{\cos\beta'}{\cos\gamma'} \mathcal{N}_{12}
(\mathcal{M}_{12}^*)^{-1}]^{-1} \mathcal{T}_{23} [\mathcal{N}_{12}
+ \mathcal{N}_{12}^{*} (\mathcal{M}_{12}^*)^{-1} \mathcal{M}_{12}]
\label{3.35} \\
&&U_F - \frac{\cot\beta'}{\cot\gamma'} \mathcal{P}_{12} =
\frac{\cot\beta'}{\cot\gamma'} \mathcal{P}_{12}^{*}V_K
\label{3.36} \\
&&U_F' - \mathcal{M}_{12} = \mathcal{M}_{12}^{*}V_{K}'
\label{3.37}
\end{eqnarray}
where we define four complex matrices
\begin{eqnarray}
\mathcal{P}_{12} &=& \frac{\cot\alpha'}{\cot\beta'} \cos(k_2 L
\cos\beta') \mathbb{I} + i \frac{c_2}{c_1} \sin(k_2 L \cos\beta')
\mathcal{T}_{12} \label{3.38} \\
\mathcal{Q}_{12} &=& \frac{c_2}{c_1} \cos(k_2 L \cos\beta')
\mathcal{T}_{12} + i \frac{\cot\alpha'}{\cot\beta'} \sin(k_2 L
\cos\beta') \mathbb{I} \label{3.39} \\
\mathcal{M}_{12} &=& \cos(k_2 L \cos\beta') \mathbb{I} + i
\frac{\cos\alpha'}{\cos\beta'} \sin(k_2 L \cos\beta')
\mathcal{T}_{12} \label{3.40} \\
\mathcal{N}_{12} &=& \frac{\cos\alpha'}{\cos\beta'} \cos(k_2 L
\cos\beta') \mathcal{T}_{12} + i \sin(k_2 L \cos\beta') \mathbb{I}
\label{3.41}
\end{eqnarray}

Then the Faraday and Kerr rotation can be given by the following
\begin{eqnarray}
\tan\theta_F &=& - \frac{1}{\sin\gamma'} \frac{E_{3z}^+}{E_{3y}^+}
= - \frac{Y_1}{2\alpha} \frac{\sin\alpha'}{\sin\gamma'}
\frac{U_F^{12}}{U_F'^{11}} = - \frac{1}{2\alpha}
\frac{\sin\alpha'}{\sin\gamma'} \frac{U_F^{12}}{U_F'^{11}}
\label{3.42} \\
\tan\theta_K &=& \frac{1}{\sin\alpha'} \frac{E_{1z}^-}{E_{1y}^-} =
\frac{Y_1}{2 \alpha} \frac{V_K^{12}}{V_K'^{11}} =
\frac{1}{2\alpha} \frac{V_K^{12}}{V_K'^{11}} \label{3.43}
\end{eqnarray}

In general, Eq.(\ref{3.42}) and (\ref{3.43}) depend on a
complicated way on the optical constants of both the TI and the
substrate, as well as on the TI film thickness L, the wave
frequency $\omega$ and the angle of incidence. Since matrices
$U_F(U_F')$ and $V_K(V_K')$ are complex, both Faraday and Kerr
angle are complex, that means that the transmitted and reflected
electric fields will acquire some ellipticity in addition to the
rotation of the plane of polarization. But in the dc limit, i.e.
$\omega \rightarrow 0$, all the values are reduced to real number,
Eq.(\ref{3.42}) and (\ref{3.43}) become
\begin{eqnarray}
\tan\theta_F &=& \frac{2 \alpha{p}}{Y_3 + \frac{\cos\gamma'}{\cos\alpha'}} \label{3.44} \\
\tan\theta_K &=& \frac{4 \alpha{p}}{Y_3^{2} - 1 + 4 \alpha^{2}
p^{2} + Y_3 (\frac{\cos\gamma'}{\cos\alpha'} -
\frac{\cos\alpha'}{\cos\gamma'})} \label{3.45}
\end{eqnarray}

So in low frequency limit, both Faraday and Kerr rotations depend
only on the optical constants of substrate, the angle of
incidence, and the important parameter $p$. All the preview work
Ref.\cite{joseph, tse1} are
focus on the case of normal incidence, which can be directed
acquired from Eqs.(\ref{3.44})-(\ref{3.45}) by letting $\alpha'
\rightarrow 0$.

\section{generalized quantization condition in TME}

From Eqs.(\ref{3.34})-(\ref{3.43}), it can be obviously seen that
$\tan\theta_F, \tan\theta_K$ are periodic functions of the
combination of $k_2L\cos\beta'$  because of the periodicity of the
matrices functions $\mathcal{P}_{12}, \mathcal{Q}_{12},
\mathcal{M}_{12}, \mathcal{N}_{12}$. More specifically, when $k_2
L \cos\beta' = n \pi, (n \in \mathbb{Z})$ , both
$\tan\theta_F,~\tan\theta_K$ are equal to the results in dc limit
separately. The condition $k_2 L \cos\beta' = n \pi$ is
equivalently to the constraint to the thickness of TI film, i.e.
\begin{eqnarray}
L \cos\beta' = n \frac{\lambda_2}{2},~~~~(n \in \mathbb{Z})
\label{4.46}
\end{eqnarray}
where $\lambda_2 = \frac{2\pi{c}}{\omega
\sqrt{\varepsilon_2\mu_2}}$ is the optical wavelength in the TI.
On the other hand, these certain values are exactly corresponding
to the minima of reflectivity. In Ref.\cite{joseph}, the author
proposed an experiment which need to tune the photon frequency
$\omega$ or the thickness of TI film $L$ to that special values in
order to observing that minima of reflectivity, but neither
$\omega$ nor $L$ can be continuously tuned by convenient ways. But
in the case of oblique incidence, we introduce an extra parameter,
$\alpha'$, which can be continuously changed conveniently in
experiments. In view of this reason, we propose a new experiment
scenario: firstly, one choose a TI film with appropriate thickness
and depose it on a topological trivial substrate, and then tune
the angle of incidence from $0$ to $\frac{\pi}{2}$ smoothly,
measure $\theta_F$ and $\theta_K$ when it occurs at reflectivity
minima. At these values, both $\tan\theta_F$ and $\tan\theta_K$
have the same expression with Eq.(\ref{3.44}) and (\ref{3.45}),
and
\begin{eqnarray}
\tan\theta_F = \frac{2\alpha{p}}{Y_3 +
\frac{\cos\gamma'}{\cos\alpha'}},~~~ \tan\theta_K =
\frac{4\alpha{p}}{Y_3^{2} - 1 + 4 \alpha^{2} p^{2} + Y_3
(\frac{\cos\gamma'}{\cos\alpha'} -
\frac{\cos\alpha'}{\cos\gamma'})}, ~~~L \cos\beta' = n
\frac{\lambda_2}{2} \label{4.47}
\end{eqnarray}
which are also dependent on the angle of incidence. For example,
if the thickness of TI film is 3.2 (with the unit of
$\frac{\lambda_2}{2}$), then we can strictly get reflectivity
minima three times altogether in the process of tuning angle of
incidence.

\begin{figure}
\includegraphics[scale=0.5]{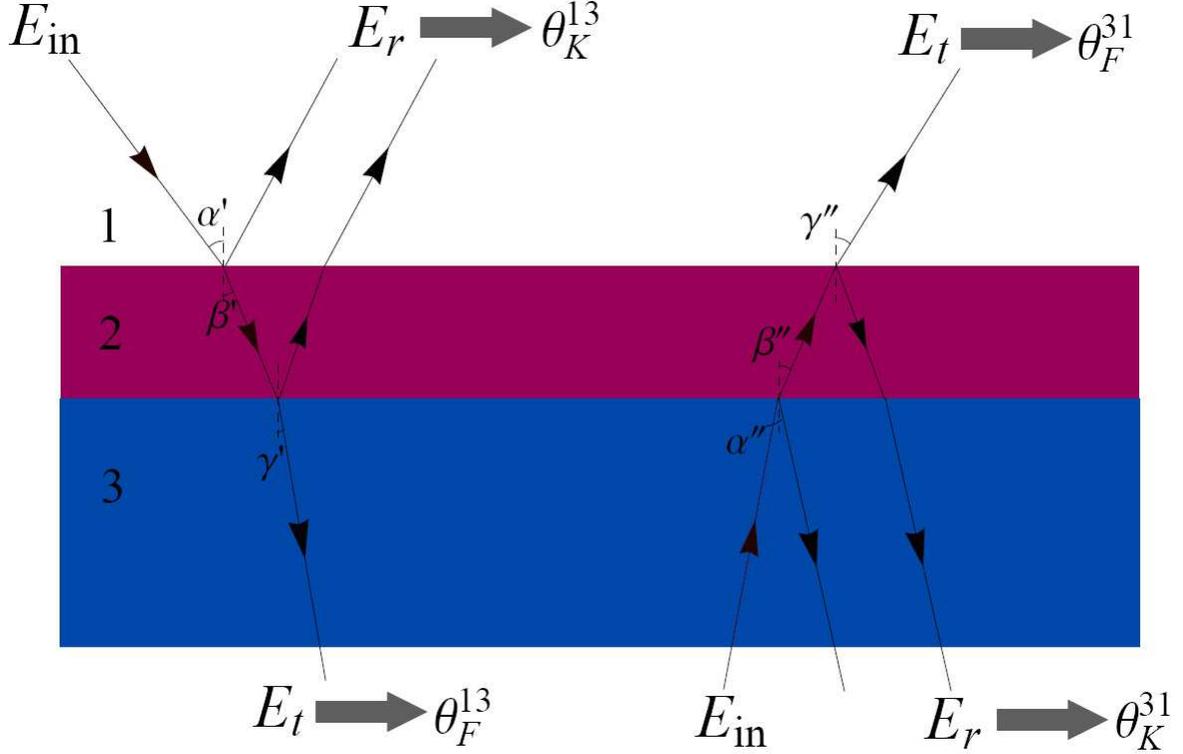}
\caption{Kerr and Faraday measurements in both directions}
\label{fig.3}
\end{figure}

In the case of orthogonal incidence, one can eliminate the
explicit dependence on the substrate property $Y_3$ by combining
$\theta_F$ and $\theta_K$  and  then obtain a topological
quantization condition in TME, but unfortunately, it does't work
in the case of oblique incidence for the emergence of an extra
parameter $\alpha'$. However, this difficulty can be solved if we
measure the Faraday and Kerr angles in both direction.  We will
elaborate this idea in the following discussion. We denote by
$\theta_F^{13}$ and $\theta_K^{13}$ the Faraday and Kerr angles
defined previously in Eq.(\ref{3.44}) and (\ref{3.45}),
respectively. While $\theta_F^{31}$ and $\theta_K^{31}$ represent
the Faraday and Kerr angles for light ray travelling in the
opposite direction shown in Fig.[\ref{fig.3}]. When light incident
from the substrate and get reflectivity minima, we obtain
\begin{eqnarray}
\tan\theta_F^{31} &=& \frac{2\alpha{p}}{1 + \frac{\cos\gamma''}{\cos\alpha''} (Y_3 + \frac{4 \alpha^2 p^2}{Y_3})} \label{4.48}\\
\tan\theta_K^{31} &=& - \frac{4 \alpha{p} Y_3}{Y_3^{2} - 1 + 4
\alpha^2 p^2 +Y_3 (\frac{\cos\alpha''}{\cos\gamma''} -
\frac{\cos\gamma''}{\cos\alpha''})} \label{4.49}
\end{eqnarray}
where denote by $\alpha'', \beta'', \gamma''$ the angle in
substrate, TI film and vacuum, separately. Eq.(\ref{4.48}) and
(\ref{4.49}) independent of TI properties but depend on the angle
of incidence. The condition of reflection minima is
$k_2L\cos\beta''=m\pi$, i.e.
\begin{eqnarray}
L \cos\beta'' = m \frac{\lambda_2}{2},~~(m \in \mathbb{Z})
\label{4.50}
\end{eqnarray}
Comparing Eqs.(\ref{4.46}) and Eqs.(\ref{4.50}), we can find that
the sets of angles, $\alpha'(\beta', \gamma')$ and
$\alpha''(\beta'', \gamma'')$ exist a one to one corresponding
relation, i.e.
\begin{eqnarray}
\beta'' = \beta',~~\alpha'' = \gamma',~~\gamma'' = \alpha',~~~(m =
n) \label{4.51}
\end{eqnarray}
substitute Eq.(\ref{4.51}) into Eq.(\ref{4.48}) and (\ref{4.49}),
we find
\begin{eqnarray}
\tan\theta_F^{31}
&=& \frac{2 \alpha{p}}{1 + \frac{\cos\alpha'}{\cos\gamma'}(Y_3 + \frac{4 \alpha^2 p^2}{Y_3})} \label{4.52} \\
\tan\theta_K^{31} &=& - \frac{4 \alpha{p} Y_3}{Y_3^{2} - 1 + 4
\alpha^2 p^2 + Y_3 (\frac{\cos\gamma'}{\cos\alpha'} -
\frac{\cos\alpha'}{\cos\gamma'})} \label{4.53}
\end{eqnarray}
comparing Eq.(\ref{3.45}) and Eq.(\ref{4.53}), we obtain that
$Y_3$ can be given as
\begin{eqnarray}
Y_3 = - \cot\theta_K^{13} \tan\theta_K^{31} \label{4.54}
\end{eqnarray}
and this relation is also true in the case of orthogonal
incidence. Moreover, the four angles, i.e. $\theta_K^{13},
~\theta_F^{13}, ~\theta_K^{31}, ~\theta_F^{31}$, can be combined
to obtain a universal result independent of both TI and substrate
properties,
\begin{eqnarray}
\frac{\cot\theta_F^{13} \tan\theta_F^{31} - \cot\theta_K^{13}
\tan\theta_K^{31}}{\cot\theta_F^{13} + \tan\theta_K^{13}
\tan\theta_F^{31} \cot\theta_K^{31}} = 2 \alpha p,~~~p \in
\mathbb{Z} \label{4.55}
\end{eqnarray}

However, it should be pointed out that Eq.(\ref{4.55}) does not
depend explicitly on the angle of incidence, but the two sets of
rotation angles in both direction must be measured at certain
condition. Specifically, if $\theta_F^{13}$ and $\theta_K^{13}$
are measured at the nth time of reflectivity minima when the angle
of incidence $\alpha'$ continuously tuned from $0$ to
$\frac{\pi}{2}$ , then $\theta_F^{31}$ and $\theta_K^{31}$ must be
measured at the nth time of reflectivity minima also, otherwise,
$\alpha'' = \gamma'$. Equation (\ref{4.55}) provides a universal
topological quantization of the TME in units of the fine structure
constant $\alpha$, independent of material properties such as
$\varepsilon$ and $\mu$, and it applies to  both normal incidence
and oblique incidence, which is the central result of our
work.

Similar to the case of normal incidence, at reflectivity minima
our generalized quantization condition depends only on $p$ and not
$\theta$. Therefore, this experiment measures the total Hall
conductance of both interfaces. The problem measuring $\theta$, or the
Hall conductances of each surfaces separately, can be solved
by the same method of Ref.\cite{joseph}. The basic idea is: to
obtain the axion angle $\theta$ we can do another optical
measurement performed at reflectivity maxima, $L \cos \beta' = (n
+ \frac{1}{2}) \frac{\lambda_2}{2},~~n \in \mathbb{Z}$, and the
rotation angle corresponding to reflectivity maxima depend
explicity on the TI axion angle $\theta$. And then using all the
measurements to construct a universal function of a single
variable $f(\theta)$ which crosses zero at the value of the bulk
axion angle $\theta$ with no $2\pi$ ambiguity. The zero crossing
point is independent of material parameters and, together with
Eq.(\ref{4.55}), provides a universal experimental demonstration
of the universal quantization of the TME in the bulk of a TI.

\section{Conclusions}

In this article, we  work out  manifestly $SL(2, \mathbb{Z})$
electric-magnetic duality covariant expression for the Kerr and
Faraday angles at oblique incidence at a single surface between a
trivial insulator and a semi-infinite topological insulator, as
well as at a topological insulator thick film with two surfaces.
When light incident at a topological insulator thick film with a
finite incidence angle, we give a generalized topological
quantization condition by combining two sets of Faraday and Kerr
angles in both direction which are all measured at reflectivity
minima. The generalized topological
quantization condition obtained here can be easier to realize
experimentally compared with an earlier proposal\cite{joseph},
since the incidence angle can be continuously tuned.

\section*{Acknowledgement}

We acknowledge helpful discussions with Liang Chen, Meng-Su Chen,
Zhong Wang, Xiaoliang Qi and Liang Sun. This work is supported by NSFC Grant
No.10675108, the Keck Foundation and the Focus Center Research Program (FCRP) Center on
Functional Engineered Nanoarchitectonics (FENA).

\end{document}